# Anomalous Nernst effect induced terahertz emission in a single ferromagnetic film

Zheng Feng[1]*, Wei Tan[1], Zuanming Jin[2]*, Yi-Jia Chen[3], Zhangfeng Zhong[4], Liang Zhang[5,7], Song Sun[1], Jin Tang[6], Yexing Jiang[2], Po-Hsun Wu[3], Jun Cheng[4], Bingfeng Miao[4,10]*, Haifeng Ding[4,10], Dacheng Wang[1], Yiming Zhu[2], Liang Guo[5], Sunmi Shin[7], Guohong Ma[8], Dazhi Hou[9,11], Ssu-Yen Huang[3]*

[1]Microsystem & Terahertz Research Center, CAEP, Chengdu 610200, P. R. China

[2]Terahertz Technology Innovation Research Institute, Terahertz Spectrum and Imaging Technology Cooperative Innovation Center, Shanghai Key Lab of Modern Optical System, University of Shanghai for Science and Technology, Shanghai 200093, P. R. China

[3]Department of Physics, National Taiwan University, Taipei 10617, Taiwan

[4]National Laboratory of Solid State Microstructures and Department of Physics, Nanjing University, Nanjing 210093, P. R. China

[5]Department of Mechanical and Energy Engineering, Southern University of Science and Technology, Shenzhen 518055, P. R. China

[6]School of Physics and Optoelectronics Engineering Science, Anhui University, Hefei 230601, P. R. China

[7]Department of Mechanical Engineering, National University of Singapore, 117516, Singapore

[8]Department of Physics, Shanghai University, Shanghai 200444, P. R. China.

[9]ICQD, Hefei National Laboratory for Physical Sciences at Microscale, University of Science and Technology of China, Hefei 230026, P. R. China

[10]Collaborative Innovation Center of Advanced Microstructures, Nanjing University, Nanjing 210093, P. R. China

[11]Department of Physics, University of Science and Technology of China, Hefei 230026, P. R. China

*fengzheng_mtrc@caep.cn, *physics_jzm@usst.edu.cn , *bfmiao@nju.edu.cn, *syhuang@phys.ntu.edu.tw

# Abstract

By developing a bidirectional-pump terahertz (THz) emission spectroscopy, we reveal an anomalous Nernst effect (ANE) induced THz emission in a single ferromagnetic film. Based on the distinctive symmetry of the THz signals, ANE is unequivocally distinguished from the previously attributed ultrafast demagnetization and anomalous Hall effect mechanisms. A quantitative method is established to separate the different contributions, demonstrating a significant ANE contribution that even overwhelms other competing mechanisms. Our work not only clarifies the origin of the ferromagnetic-based THz emission, but also offers a fertile platform for investigating the ultrafast magnetism and THz spintronics.

The terahertz (THz) wave refers to the electromagnetic radiation within the spectral window between the electronics and optics [1,2]. This region is where many fundamental collective excitations or ultrafast charge/phonon/spin dynamics occur. THz spectroscopy thus can serve as a sensitive probe to the physical insights of various materials [3-5]. Besides the importance in fundamentals, THz wave also exhibits great potential in many applications, such as communications, spectroscopy, imaging, biomedical, security, etc [6-9]. In parallel, the rich excitations in many different materials also yield a broad platform to explore novel THz devices [10-18]. One of the recent and fascinating examples is the magnetic heterostructure-based spintronic THz emitter [10,11,19-25], which provides the advantages such as ultrabroad bandwidth, high efficiency, high flexibility, and controllable polarization. As a result, the spintronic THz emitter has been successfully employed in various applications [26-28], and THz spintronics has emerged as a new and vivid research field [29-31].

A spintronic THz emitter typically consists of a ferromagnetic|nonmagnetic (FM|NM) bilayer nanofilm, where the femtosecond (fs) laser induced ultrafast spin current generation in FM layer and the ultrafast spin-charge conversion in NM layer lead to the THz pulse emission [10]. Interestingly, as the spin current source of a spintronic THz emitter, a single FM film under the fs laser pump can also generate THz pulses [32-44], which was first discovered in 2004 [32], even earlier than the invention of the spintronic THz emitter. The authors attributed the THz emission to the magnetic dipole radiation induced by the well-known ultrafast demagnetization [45-49], which was widely adopted as the exclusive mechanism in the past two decades. The THz signals originating from the ultrafast demagnetization is expected to maintain its sign when the sample is flipped but with the same magnetization orientation [35,38]. Q. Zhang et al., however, observed a sign reversal and they explained it with the anomalous Hall effect (AHE) [50-52]. Despite the fact that the THz emission from a single FM film is quite important for the study of ultrafast spin dynamics/ultrafast spin-transport [35,39,41,44], the reveal of novel spin/orbit effect [18,40,43,53], and the

development of high performance THz emitter [52], its underlying mechanism seems still not well understood.

When tracing back to the various fundamental processes, we note that one critical piece is missing in the physical picture—the laser heating effect [54-56]. Based on the facts that (i) A fs laser pulse can cause a transient temperature gradient of electrons at sub-picosecond time scale [45,54,55,57], (ii) a temperature gradient in a single FM film can induce a spin polarization charge current perpendicular the temperature gradient and magnetization due to anomalous Nernst effect (ANE) [58,59], and (iii) a time varying charge current radiates electromagnetic wave, we unequivocally speculate that the ANE can in principle generate THz emission. Very recently, different groups have observed the ultrafast spin Seebeck effect and photothermoelectric effect induced THz emission due to sub-picosecond temperature gradient in YIG|NM bilayer and Dirac semimetallic $Cd_3As_2$, respectively [60,61]. These studies show the possible essential contribution of the ultrafast thermal effect in THz emission, which, in contrast, has been overlooked in single FM films. In fact, it is very challenging to reveal and quantitatively identify the contribution of the ANE on the induced THz emission in FM films, since it is always mixed with the previously attributed ultrafast demagnetization and/or AHE induced contributions. And, as will be detailed below, the current method based on flipping sample [35,38,50-52] is insufficient to distinguish the ANE induced THz emission from the other two contributions.

Here, we revisit the THz emission mechanism in a single FM film utilizing a newly developed bidirectional pump-THz emission spectroscopy. We investigate the THz emissions of Fe films with various thicknesses in four different configurations, namely, with opposite pump directions and flipped sample sides. Surprisingly, within a certain thickness range, the pump direction and film side dependent symmetries of the observed THz signals show distinct feature which is significantly different from those generated by the ultrafast demagnetization or AHE. Evidence from diverse perspectives clearly verifies that it results from the ultrafast ANE. We further develop a quantitative method to separate THz signals from various contributions and find that the ANE induced mechanism is dominant within a certain thickness range. This work not only demonstrates the ANE induced THz emission, unifies the seemed contradictory observations in the previous studies, but also set a benchmark for the investigation of THz emission induced by different mechanisms.

## Results

### Distinctive features of spintronic THz emission via different mechanisms

We first discuss the distinctive features of the THz emission from previously reported mechanisms when both sample and laser pump direction can be flipped, including those generated by the inverse spin Hall effect (ISHE) in FM|NM bilayer, and ultrafast demagnetization and AHE in a single FM film. In a FM|NM bilayer, the THz emission is originated from the conversion of a fs laser induced ultrafast spin current $J_s(t)$ to a transient charge current $J_c(t)$ via ISHE, which has the electric field [10,11]:

$$E_{\mathrm{ISHE}}(t) \propto \frac{\partial}{\partial t}\left[J_{\mathrm{c}}(t)\right] = \frac{\partial}{\partial t}\left[\theta_{\mathrm{ISHE}} \cdot J_{\mathrm{s}}(t) \times \hat{M}\right], \quad (1)$$

where $t$ is the time, $\hat{M}$ is the unit vector of the magnetization $\vec{M}$, $\theta_{\mathrm{ISHE}}$ is the spin Hall angle of the NM layer . Equation (1) shows that the THz polarity should be reversed upon flipping the sample, namely, from FM|Sub to Sub|FM (Sub stands for substrate) as it is naturally accompanied by a reversal of $J_s(t)$, but maintains the same polarity when the laser pump direction changes, as illustrated in Fig. 1a. For the AHE induced THz emission in a single FM film, it was attributed to the conversion of a laser generated net backflow longitudinal current $J_{c'}(t)$ to a transient transverse current $J_c(t)$ (with anomalous Hall angle $\theta_{\mathrm{AHE}}$), thereby emits a THz pulse with the electric field [50] :

$$E_{\mathrm{AHE}}(t) \propto \frac{\partial}{\partial t}\left[J_{\mathrm{c}}(t)\right] = \frac{\partial}{\partial t}\left[\theta_{\mathrm{AHE}} \cdot J_{c'}(t) \times \hat{M}\right] . \quad (2)$$

As $J_{c'}(t)$ is determined by the two interfaces of the FM film, it can be easily derived that the AHE induced THz emission possess the same features of the ISHE induced THz emission, as illustrated in Fig. 1a. In contrast, the ultrafast demagnetization induced THz emission has the electric field [32]:

$$E_{\text{dem}}(t) \propto \frac{\partial^2 M(t)}{\partial t^2} \cdot \hat{M} \times \hat{z}\,, \tag{3}$$

where $M(t)$ is the time-dependent magnetization, $\hat{z}$ is the unit vector along the direction of the film normal. From Eq. (3), we find that its sign is independent with the film side and the pump direction, as illustrated in Fig. 1b.

Now we investigate the ANE induced THz emission in details. A transient spin polarized charge current $J_c(t)$ can be induced by a sub-picosecond electron temperature gradient $\nabla T(t)$ via ANE, which leads to the THz emission with the electrical field [62]:

$$E_{\text{ANE}}(t) \propto \frac{\partial}{\partial t}\left[J_c(t)\right] = \frac{\partial}{\partial t}\left[\sigma_{\text{F}} \cdot \theta_{\text{ANE}} \cdot S_{\text{F}} \cdot \hat{M} \times \nabla T(t)\right], \tag{4}$$

where $\theta_{\text{ANE}}$ is the anomalous Nernst angle, $S_{\text{F}}$ and $\sigma_{\text{F}}$ are the Seebeck coefficient and electric conductivity of FM in the film plane, respectively. Equation (4) shows that the THz polarity is determined by the sign of $\theta_{\text{ANE}}$, $S_{\text{F}}$, and the direction of $\nabla T(t)$. Note that the laser-induced temperature gradient $\nabla T$ strongly depends on the film thickness [58]. Generally, for a film thicker than a critical thickness, shorted as “thick” film, $\nabla T$ points to the pump direction of laser and is irrelevant to the sample flipping. The critical thickness is highly related to the optical penetration depth and thermal properties of the sample. On the contrary, for a film thinner than the critical thickness, shorted as “thin” film, $\nabla T$ is always pointing away from the FM|substrate interface regardless of the laser pump direction when the substrate is optically transparent with high thermal conductivity (for instance, the MgO substrate in this work). This has been demonstrated recently with the continuous wave (CW) laser [58]. The ultrafast temperature analysis and simulation (see Section S1, Fig. S1 and Fig. S2 in Supplementary Information) show that $\nabla T$ in “thick” and “thin” films at the sub-picosecond time scale are very similar with the *dc* case. The thickness dependence of ultrafast electron temperature gradient $\nabla T$ is a competition result of the “bulk” temperature gradient $\nabla T_B$ and “interface” temperature gradient $\nabla T_I$ , as illustrated in Supplementary Fig. S1. The “bulk” temperature gradient $\nabla T_B$ is determined by the laser energy absorption profile in the metal, which points to the pump direction of laser. And the “interface” temperature gradient $\nabla T_I$ is caused by the heat transfer from the metal to the dielectric substrate through the interface due to the coupling between metal electrons and substrate phonons (coupling time could be shorter than 1ps) [63], which points away from the FM|substrate interface. The “bulk” temperature gradient is dominant in “thick” films, while the “interface” temperature gradient is dominant in “thin” films. Therefore, the feature of ANE induced THz emission will be different for “thick” and “thin” FM films, as shown in Fig. 1c. For a “thick” FM film, the THz polarity remains the same upon the sample flipping, but flips its sign when the pump direction is reversed. Importantly, this feature is clearly different from those discussed above and can serve as a fingerprint to distinguish ANE from the other mechanisms. For a “thin” FM film, it shows the same feature as the AHE induced THz emission, making it difficult to distinguish them. In such case, additional evidence is needed as will be discussed below.

**Bidirectional pump-THz emission spectroscopy**

To implant the above analysis into practice, we develop a bidirectional pump-THz emission spectroscopy, where both the pump direction and the sample orientation can be readily selected. The diagram is displayed in Fig. 2. The femtosecond laser pulse (with duration of 100 fs, center wavelength of 800 nm, average power of 2W, and repetition rate of 80 MHz) is first split into two beams acting as pump and probe beams, respectively. The pump beam is further split into forward and backward pump beams, which pass through optical delay line 1 (ODL 1) and the hole of the off-axis parabolic mirror 1 (OAPM 1), respectively. The forward and backward pump beams illuminate on the same position of the sample with the same spot size (diameter: ~0.2mm).

An in-plane magnetic field of 80 mT was applied to the sample, and the resultant THz electric fields, which are perpendicular to the magnetic fields, are collected and detected by the electro-optic sampling with probe beam. By carefully adjusting the ODL 1 to obtain identical optical path of the forward and backward pump beams, the THz pulses induced by the two beams appear at same time. In the sampling detection, 2-mm-thick ZnTe crystal is used. This design supports the operation of the energy flow reversing, in addition to the operation of sample flipping. (Note that for the operation of sample flipping, it is necessary to maintain the film position unchanged.) It's worth mentioning that the sample flipping dependence was also utilized by previous studies [35,38,50-52], which, however, is still insufficient to identify different THz emission mechanisms as shown in Fig. 1. As will be further discussed below, our bidirectional pump setup overcomes the shortage and provides a critical probe.

**Verification of the ANE mechanism**

Fe(2 nm)|Pt(2 nm) bilayer is chosen to test the validity of the proposed method. As shown in Fig. 3a, the THz signal show the same sign with either the forward or backward pump, and reverses its sign upon the sample flipping. This behavior is consistent with the feature of the ISHE induced THz emission illustrated in Fig. 1a.

We continue to measure the THz emission for single Fe films with the thickness varying from 4 nm to 100 nm. Figure 3b shows the THz signal of Fe(100 nm) with four different configurations. In contrast to the case of Fe(2 nm)|Pt(2 nm), it maintains its polarity in all these conditions. This feature agrees well with that of the ultrafast demagnetization induced THz emission as shown in Fig. 1b, suggesting it is the dominant contribution. Figure 3c depicts the THz signals of Fe(55 nm) whose thickness is much larger than the optical penetration depth of Fe, which is ~17.3 nm at the wavelength of 800 nm [64]. It reverses its sign with opposite pump directions, but maintains its sign upon flipping the sample. This feature is clearly different from that of the demagnetization or the AHE induced THz emission, and can't be explained by the combination of both, either. Instead, it coincides with the feature of the ANE induced THz emission in “thick” films, as illustrated in Fig. 1c. We also found that the THz emission shows the same feature within the thickness range of 45-90 nm (see Supplementary Fig. S3). Although the lineshapes referring to different thicknesses exhibit certain differences, the four measured signals of a specific Fe film are quite similar and the polarity can be intuitively distinguished. This finding unambiguously demonstrates the ANE induced THz emission within this thickness range.

Moreover, when the thickness of the Fe film further decreases, we find another change of the THz signal symmetry. As shown in Fig. 3d, the THz signal of Fe(9 nm) reverses its sign upon the sample flipping, while remaining its sign with reversed pump directions. At the first glance, it agrees well with the picture of ANE in a “thin” film illustrated in Fig. 1d. The THz signals of Fe(19, 15, 5, and 4 nm) also show similar feature (see Supplementary Fig. S3). Nevertheless, we still cannot determine whether it is from ANE or AHE since they have identical symmetry (see Fig. 1). Hence, the next question is how to identify ANE and AHE mechanisms in “thin” films. We notice that the ANE of Fe film changes its sign between 4 nm and 5 nm [65], whereas the AHE of Fe film remains the same sign from 1.5 to 93 nm [66]. Thus, inspecting the sign change near the critical thickness provides a promising method to distinguish these two mechanisms. Figures 4a and 4b show the measured THz signals of Fe(4 nm) and Fe(5 nm), respectively. They apparently show the opposite signs. We further measured their *dc* ANE and AHE responses. As shown in Figs. 4c-4f, the ANE voltages show opposite signs, in line with the observed sign change of the THz signals, whereas the AHE voltages present the same sign. The comparison among the polarities of THz emission and the ANE/AHE voltages clearly demonstrates that the ANE dominates THz emission also in the “thin” FM films. Meanwhile, further experiments show the laser polarization independence of the THz emission of Fe films (see Section S3 in Supplementary Information), which excludes the contribution of the optical rectification effects [67,68].

**Quantitative separation of multiple mechanisms based on symmetry analysis**

Above, we exhibited that the ANE dominates the THz emission in single FM films within a certain thickness range. In the following, we discuss a quantitative and self-consistent method to separate various contributions according to their symmetries shown in Fig. 1. The separation method is based on that the total THz emission is the superposition of emission of the ultrafast demagnetization and the ANE induced mechanisms. (Note that the AHE is quite weak in the above measurements thus being neglected). Meanwhile, the ANE contribution is the superposition of the two contributions induced by the "bulk" and "interface" temperature gradients [58], where the "bulk" component and "interface" component present the ANE features of "thick" film and "thin" film, respectively. For the details of the separation, please see Methods. Figure 5 summarizes the main results. The thickness dependent of the demagnetization contribution in FM|Sub condition under both forward and backward pump, Fig. 5a, show the positive THz signals which firstly increase with increasing Fe thickness and eventually saturate. This is in line with the theoretical trends that have been deduced and discussed by T. J. Huisman et al. [35]. The slightly difference in amplitude may originate from the different absorption in forward/backward pump. Figure 5b shows the thickness dependent of the separated ANE contribution under the same condition. Importantly, the ANE contribution is significant and far outweighs the demagnetization contribution in thin film region, and the demagnetization mechanism will dominate in the thick film region (>90 nm). It is interesting to note that, below 20 nm, both forward and backward pump of ANE component exhibit the same sign, whereas above 40 nm, the signals with two pump configurations show similar amplitudes but with opposite signs. These coincide with the thickness dependences of the electron temperature gradient (See Supplementary Fig. S1).

## Discussion

In the above, we transparently distinguished the different THz emission mechanisms coexisted in a single Fe film. It is worth noting that although the ultrafast demagnetization only dominates within the thickness range of >90 nm, it contributes apparently with all thickness, which is consistent with the widely accepted picture. However, in previous studies with only sample flipping, it shows the same symmetry with the "bulk" ANE, which made the ANE overlooked. By further employing the operation of the energy flow reversing in this work, the ANE contribution was unveiled directly and confirmed as the dominant mechanisms within the thickness range of ≤90 nm. In addition, the seemed contradictory observations in previous studies—the THz polarity remains upon sample flipping in some studies [35,38] but reverses in others [50-52]—can also be unified within our framework. The fact lies on that the competition of several coexisting spintronic processes determines the dominant feature of the THz emission, for example, the ultrafast demagnetization, the "bulk" ANE, and the "interface" ANE. We elucidated that the dominant mechanism changes with varying thickness, and it is highly expected to alter with distinct FM materials and substrate with different thermal properties.

We also note that since the AHE owns the same symmetry feature with the "interface" ANE, they can hardly be separated by only employing the above symmetry analysis. We raised an exclusive method by inspecting and comparing the sign of THz/ANE/AHE signals. Specifically in this work, Fe(4 nm) and Fe(5 nm) films, which have same-sign AHE but opposite-sign ANE, emit THz wave with opposite polarization, demonstrating that the "interface" ANE is much larger than the AHE in Fe films. On the other hand, it can be imagined that in other FM materials with smaller ANE and/or with giant difference of electron scattering at two interfaces, the AHE contribution may has non-ignorable value. Hence, how to directly separate the AHE contribution from the ANE contribution within a universal framework is still an open question, which needs further exploration.

In summary, by developing a home-built bidirectional pump-THz emission spectroscopy and the associated symmetry analysis, we have demonstrated the ANE induced THz emission in single Fe films unambiguously. Based on the distinct symmetry features of different mechanisms, we developed a quantitative method to separate the different contributions in THz emission. In sharp contrast with the previously attributed mechanisms of ultrafast demagnetization and AHE, the contribution of the THz emission induced by the ANE compellingly dominates in thin Fe films. This work deepens the understanding of the THz emission of single magnetic films and could promote the revisit of the mechanisms of spintronic THz emitters, since the thermal effect was overlooked previously. Moreover, it also provides a fertile platform for the development of new type of THz emitter either from the ultrafast spin caloritronics processes alone, or in combination with other non-thermal mechanisms.

# Methods

## Sample preparations

The Fe(2nm) |Pt(2nm) bilayer and the Fe single films with different thicknesses were grown at 300 K on MgO(001) substrate by magnetron sputtering. All the Fe films were covered by 5nm thick $SiO_2$ layer for protection. As the thermal conductivity of $SiO_2$ is very small, the $SiO_2$ capping layers have negligible influence on the electron temperature gradient of Fe films.

## Ultrafast temperature simulation

The ultrafast temperature simulation is based on the two-temperature model (TTM) involving the heat transfer across the metal-dielectric interface [69]. The spatial and temporal evolution of the electron and phonon temperatures can be described by:

$$C_e \frac{\partial T_e}{\partial t} = k_e \frac{\partial^2 T_e}{\partial z^2} - G(T_e - T_p) + S(z,t)\,, \tag{5}$$

$$C_p \frac{\partial T_p}{\partial t} = k_p \frac{\partial^2 T_p}{\partial z^2} + G(T_e - T_p)\,, \tag{6}$$

$$C_c \frac{\partial T_c}{\partial t} = k_c \frac{\partial^2 T_c}{\partial z^2}, \tag{7}$$

$$C_s \frac{\partial T_s}{\partial t} = k_s \frac{\partial^2 T_s}{\partial z^2}, \tag{8}$$

with the initial conditions:

$$T_e(t=0) = T_p(t=0) = T_c(t=0) = T_s(t=0) = T_0, \tag{9}$$

and the interface conditions:

$$-k_e \frac{\partial T_e}{\partial z}\bigg|_{z=0} = -\frac{T_e - T_c}{R_{ec}}\bigg|_{z=0}, \; -k_e \frac{\partial T_e}{\partial z}\bigg|_{z=d} = -\frac{T_s - T_e}{R_{es}}\bigg|_{z=d}, \tag{10}$$

$$-k_p \frac{\partial T_p}{\partial z}\bigg|_{z=0} = -\frac{T_p - T_c}{R_{pc}}\bigg|_{z=0}, \; -k_p \frac{\partial T_p}{\partial z}\bigg|_{z=d} = -\frac{T_s - T_p}{R_{ps}}\bigg|_{z=d}, \tag{11}$$

$$-k_c \frac{\partial T_c}{\partial z}\bigg|_{z=-d_1} = 0, \; -k_c \frac{\partial T_c}{\partial z}\bigg|_{z=0} = -\frac{T_e - T_c}{R_{ec}}\bigg|_{z=d} - \frac{T_p - T_c}{R_{pc}}\bigg|_{z=d}, \tag{12}$$

$$-k_s \frac{\partial T_s}{\partial z}\bigg|_{z=d} = -\frac{T_s - T_e}{R_{es}}\bigg|_{z=d} - \frac{T_s - T_p}{R_{ps}}\bigg|_{z=d}, \; -k_s \frac{\partial T_s}{\partial z}\bigg|_{z=d+d_2} = 0, \tag{13}$$

where the subscripts *e* and *p* denote electrons and phonons in the metal layer (Fe), respectively, and the subscripts *c* and *s* denote phonons in the capping layer ($SiO_2$) and the substrate (MgO), respectively. *C* is the volumetric heat capacity, *k* is the thermal conductivity, *G* is the electron–phonon coupling factor governing the energy transfer strength from electrons to phonons in metal, *S* is the fs laser heating source, and $T_0$ is the initial temperature (300K). In the interface conditions, $-k\frac{\partial T}{\partial z}$ shows the heat flux across the interface, and *R* is the interface thermal resistance [63,69]. $R_{ec}$ ($R_{pc}$) indicates the coupling strength between electrons (phonons) in the metal layer and phonons in the capping layer, while $R_{es}$ ($R_{ps}$) indicates the coupling strength between electrons (phonons) in metal and phonons in substrate. (Large thermal resistance corresponds to weak coupling and small heat flux.) $d$, $d_1$ and $d_2$ are the thickness of the metal film, capping layer, and substrate, respectively.

The fs laser heating source term $S$ is represented as

$$S = 0.94 \frac{(1 - R_{fs} - T_{fs})F}{\delta t_p (1 - \exp(-\frac{d}{\delta}))} \exp\left[ -\frac{z}{\delta} - 2.77 \left( \frac{t}{t_p} \right)^2 \right], \tag{14}$$

where $t_p$ is the full width at half maximum(FWHM) of the Gaussian type fs laser pulse, $F$ is the fluence of the laser, and $\delta$ is the optical penetration depth of metal. $R_{fs}$ and $T_{fs}$ is the reflectance and transmission of the fs laser, which could be measured.

With the thermophysical parameters listed in Supplementary Table S1, we perform ultrafast temperature calculation in MATLAB, and the calculated results are shown in Supplementary Fig. S2.

**Separation method of THz emission from ANE and ultrafast demagnetization**

The separation method is based on the fact that the THz emission in a single FM film is mainly the superposition of the emission induced by the ultrafast demagnetization and ANE, and the ANE contribution is the superposition of the two contributions induced by the "bulk" and "interface" temperature gradients. The THz emission induced by ultrafast demagnetization, the "bulk" temperature gradients $\nabla T_B$ and the "interface" temperature gradients $\nabla T_I$ are denoted by $D$, $B$, and $I$ respectively. For simplicity, we assume the identical laser absorption and identical THz radiation impendence Z for the four different configurations. Then as illustrated in Fig. 6a, the THz emission of Fe|Sub upon the forward pump ($E_{F\text{-}F}$) and the backward pump ($E_{F\text{-}B}$) can be described as:

$$E_{F-F} = D + B + I, \tag{15}$$

$$E_{F-B} = D - B + I. \tag{16}$$

As illustrated in Fig. 6b, the THz emission of Sub|Fe upon the forward pump ($E_{S\text{-}F}$ ) and the backward pump ($E_{S\text{-}B}$) can be written as:

$$E_{S-F} = D + B - I \,, \tag{17}$$

$$E_{S-B} = D - B - I \,. \tag{18}$$

From above four equations, we can obtain the THz emission induced by ultrafast demagnetization, and the ANE of “interface” and “bulk” temperature gradient respectively:

$$D = (E_{F-F} + E_{S-B})/2 \quad \text{or} \quad D = (E_{F-B} + E_{S-F})/2 \,, \tag{19}$$

$$B = (E_{F-F} - E_{F-B})/2 \quad \text{or} \quad B = (E_{S-F} - E_{S-B})/2 \,, \tag{20}$$

$$I = (E_{F-F} - E_{S-F})/2 \quad \text{or} \quad I = (E_{F-B} - E_{S-B})/2 \,, \tag{21}$$

In real operation, the laser absorption and THz radiation impendence should be considered. Assuming $E_{F\text{-}F}$ is equal to its original THz waveform (used as the normalized waveform), then $E_{F\text{-}B}$ is obtained by multiplying its original THz waveform by a factor $f_{abs}$ ( the ratio of laser absorption of incidence on FM side and Sub side), $E_{S\text{-}B}$ is obtained by multiplying its original THz waveform by a factor $f_Z$ (the ratio of THz radiation impendence of radiation out from Sub side and FM side), and $E_{S\text{-}F}$ is obtained by multiplying its original THz waveform by the factors $f_{abs}$ and $f_Z$; $f_{abs}$ is measured, $f_Z$ is equal to $2n/(1+n)$ where $n$ is the refractive index of substrate (=3.1 for MgO). After such normalization, the two THz waveforms of Sub|Fe are shifted to the same time position of Fe|Sub, as illustrated in Supplementary Fig. S5(a1-e1), where the Fe thickness are chosen with five typical thicknesses (100 nm, 85 nm, 55 nm, 40 nm and 9 nm). Then $D$, $B$, and $I$ were calculated from the four THz waveforms according to Eqs. (19-21). We note that every component can be obtained by two different calculations individually. The results obtained with these two independent methods are almost identical for every component with the same Fe thickness and this one-to-one correspondence can be found with almost all Fe thicknesses, as shown in Supplementary Fig. S5. This confirms the validity of our separation method and evidences that our method is self-consistent. At last, we obtain the Fe thickness dependent of the amplitude of the demagnetization induced THz emission under forward pump ($D$) and backward pump ($D/f_{abs}$), and the amplitude of ANE induced THz emission under forward

pump($B$+$I$) and backward pump (($-B$+$I$)/$f_{abs}$). The results are summarized in Figs. 5a and 5b. There, $D$, $B$, and $I$ are the average value of the two calculations based on Eqs. (19-21).

## Data availability

All data will be readily available upon request. Expert assistance and guide to the pertinent data will be provided for the best interest of the researchers who wish to use our data for their prospective research.

## References

1 Ferguson, B. & Zhang, X.-C. Materials for terahertz science and technology. *Nat. Mater.* **1**, 26-33 (2002).

2 Tonouchi, M. Cutting-edge terahertz technology. *Nat. Photonics* **1**, 97-105 (2007).

3 Basov, D. N., Averitt, R. D., van der Marel, D., Dressel, M. & Haule, K. Electrodynamics of correlated electron materials. *Rev. Mod. Phys.* **83**, 471-541 (2011).

4 Ulbricht, R., Hendry, E., Shan, J., Heinz, T. F. & Bonn, M. Carrier dynamics in semiconductors studied with time-resolved terahertz spectroscopy. *Rev. Mod. Phys.* **83**, 543-586 (2011).

5 Kirilyuk, A., Kimel, A. V. & Rasing, T. Ultrafast optical manipulation of magnetic order. *Rev. Mod. Phys.* **82**, 2731-2784 (2010).

6 Federici, J. & Moeller, L. Review of terahertz and subterahertz wireless communications. *J. Appl. Phys.* **107**, 111101 (2010).

7 Jepsen, P. U., Cooke, D. G. & Koch, M. Terahertz spectroscopy and imaging – Modern techniques and applications. *Laser Photonics Rev.* **5**, 124-166 (2011).

8 Peng, Y., Shi, C., Zhu, Y., Gu, M. & Zhuang, S. Terahertz spectroscopy in biomedical field: a review on signal-to-noise ratio improvement. *PhotoniX* **1**, 12 (2020).

9 Davies, A. G., Burnett, A. D., Fan, W., Linfield, E. H. & Cunningham, J. E. Terahertz spectroscopy of explosives and drugs. *Mater. Today* **11**, 18-26 (2008).

10 Kampfrath, T. *et al.* Terahertz spin current pulses controlled by magnetic heterostructures. *Nat. Nanotechnol.* **8**, 256-260 (2013).

11 Seifert, T. *et al.* Efficient metallic spintronic emitters of ultrabroadband terahertz radiation. *Nat. Photonics* **10**, 483-488 (2016).

12 Cheng, R., Xiao, D. & Brataas, A. Terahertz Antiferromagnetic Spin Hall Nano-Oscillator. *Phys. Rev. Lett.* **116**, 207603 (2016).

13 Li, J. *et al.* Spin current from sub-terahertz-generated antiferromagnetic magnons. *Nature* **578**, 70-74 (2020).

14 Vaidya, P. *et al.* Subterahertz spin pumping from an insulating antiferromagnet. *Science* **368**, 160-165 (2020).

15 Gomonay, O., Jungwirth, T. & Sinova, J. Narrow-band tunable terahertz detector in antiferromagnets via staggered-field and antidamping torques. *Phys. Rev. B* **98**, 104430 (2018).

16 Welp, U., Kadowaki, K. & Kleiner, R. Superconducting emitters of THz radiation. *Nat. Photonics* **7**, 702-710 (2013).

17 Wang, X. B. *et al.* Topological-insulator-based terahertz modulator. *Sci. Rep.* **7**, 13486 (2017).

18 Wang, X. *et al.* Ultrafast Spin-to-Charge Conversion at the Surface of Topological Insulator Thin Films. *Adv. Mater.* **30**, 1802356 (2018).

19 Yang, D. *et al.* Powerful and Tunable THz Emitters Based on the Fe/Pt Magnetic Heterostructure. *Adv. Opt. Mater.* **4**, 1944-1949 (2016).

20 Wu, Y. *et al.* High-Performance THz Emitters Based on Ferromagnetic/Nonmagnetic Heterostructures. *Adv. Mater.* **29**, 1603031 (2017).

21 Torosyan, G., Keller, S., Scheuer, L., Beigang, R. & Papaioannou, E. T. Optimized Spintronic Terahertz Emitters Based on Epitaxial Grown Fe/Pt Layer Structures. *Sci. Rep.* **8**, 1311 (2018).

22 Feng, Z. *et al.* Highly Efficient Spintronic Terahertz Emitter Enabled by Metal–Dielectric Photonic Crystal. *Adv. Opt. Mater.* **6**, 1800965 (2018).

23 Feng, Z. *et al.* Spintronic terahertz emitter. *J. Appl. Phys.* **129**, 010901 (2021).

24 Seifert, T. S., Cheng, L., Wei, Z., Kampfrath, T. & Qi, J. Spintronic sources of ultrashort terahertz electromagnetic pulses. *Appl. Phys. Lett.* **120**, 180401 (2022).

25 Jin, Z. *et al.* Cascaded Amplification and Manipulation of Terahertz Emission by Flexible Spintronic Heterostructures. *Laser Photonics Rev.* **16**, 2100688 (2022).

26 Chen, S.-C. *et al.* Ghost spintronic THz-emitter-array microscope. *Light Sci. Appl.* **9**, 99 (2020).

27 Bulgarevich, D. S. *et al.* Terahertz Magneto-Optic Sensor/Imager. *Sci. Rep.* **10**, 1158 (2020).

28 Liu, S. *et al.* Modulated terahertz generation in femtosecond laser plasma filaments by high-field spintronic terahertz pulses. *Appl. Phys. Lett.* **120**, 172404 (2022).

29 Walowski, J. & Münzenberg, M. Perspective: Ultrafast magnetism and THz spintronics. *J. Appl. Phys.* **120**, 140901 (2016).

30 Huisman, T. J. & Rasing, T. THz Emission Spectroscopy for THz Spintronics. *J. Phys. Soc. Jpn.* **86**, 011009 (2016).

31 Tang, J. *et al.* Ultrafast Photoinduced Multimode Antiferromagnetic Spin Dynamics in Exchange-Coupled Fe/R$FeO_3$ (R = Er or Dy) Heterostructures. *Adv. Mater.* **30**, 1706439 (2018).

32 Beaurepaire, E. *et al.* Coherent terahertz emission from ferromagnetic films excited by femtosecond laser pulses. *Appl. Phys. Lett.* **84**, 3465-3467 (2004).

33 Meserole, C. A. *et al.* Growth of thin Fe(001) films for terahertz emission experiments. *Appl. Surf. Sci.* **253**, 6992-7003 (2007).

34 Shen, J., Zhang, H. W. & Li, Y. X. Terahertz Emission of Ferromagnetic Ni-Fe Thin Films Excited by Ultrafast Laser Pulses. *Chin. Phys. Lett.* **29**, 067502 (2012).

35 Huisman, T. J., Mikhaylovskiy, R. V., Tsukamoto, A., Rasing, T. & Kimel, A. V. Simultaneous measurements of terahertz emission and magneto-optical Kerr effect for resolving ultrafast laser-induced demagnetization dynamics. *Phys. Rev. B* **92**, 104419 (2015).

36 Kumar, N., Hendrikx, R. W. A., Adam, A. J. L. & Planken, P. C. M. Thickness dependent terahertz emission from cobalt thin films. *Opt. Express* **23**, 14252-14262 (2015).

37 Venkatesh, M., Ramakanth, S., Chaudhary, A. K. & Raju, K. C. J. Study of terahertz emission from nickel (Ni) films of different thicknesses using ultrafast laser pulses. *Opt. Mater. Express* **6**, 2342-2350 (2016).

38 Zhang, S. *et al.* Photoinduced terahertz radiation and negative conductivity dynamics in Heusler alloy $Co_2MnSn$ film. *Opt. Lett.* **42**, 3080-3083 (2017).

39 Huang, L. *et al.* Direct observation of terahertz emission from ultrafast spin dynamics in thick ferromagnetic films. *Appl. Phys. Lett.* **115**, 142404 (2019).

40 Cheng, L. *et al.* Far out-of-equilibrium spin populations trigger giant spin injection into atomically thin $MoS_2$. *Nat. Phys.* **15**, 347-351 (2019).

41 Zhang, W. *et al.* Ultrafast terahertz magnetometry. *Nat. Commun.* **11**, 4247 (2020).

42 Huang, L. *et al.* Universal field-tunable terahertz emission by ultrafast photoinduced demagnetization in Fe, Ni, and Co ferromagnetic films. *Sci. Rep.* **10**, 15843 (2020).

43 Lee, K. *et al.* Superluminal-like magnon propagation in antiferromagnetic NiO at nanoscale distances. *Nat. Nanotechnol.* **16**, 1337-1341 (2021).

44 Li, G. *et al.* Ultrafast kinetics of the antiferromagnetic-ferromagnetic phase transition in FeRh. *Nat. Commun.* **13**, 2998 (2022).

45 Beaurepaire, E., Merle, J. C., Daunois, A. & Bigot, J. Y. Ultrafast Spin Dynamics in Ferromagnetic Nickel. *Phys. Rev. Lett.* **76**, 4250-4253 (1996).

46 Bigot, J.-Y., Vomir, M. & Beaurepaire, E. Coherent ultrafast magnetism induced by femtosecond laser pulses. *Nat. Phys.* **5**, 515-520 (2009).

47 Koopmans, B. *et al.* Explaining the paradoxical diversity of ultrafast laser-induced demagnetization. *Nat. Mater.* **9**, 259-265 (2010).

48 Fähnle, M. Review of Ultrafast Demagnetization After Femtosecond Laser Pulses: A Complex Interaction of Light with Quantum Matter. *Am. J. of Mod. Phys.* **7**, 68 (2018).

49 Tengdin, P. *et al.* Critical behavior within 20 fs drives the out-of-equilibrium laser-induced magnetic phase transition in nickel. *Sci. Adv.* **4**, eaap9744 (2018).

50 Zhang, Q. *et al.* Terahertz Emission from Anomalous Hall Effect in a Single-Layer Ferromagnet. *Phys. Rev. Appl.* **12**, 054027 (2019).

51 Liu, Y. *et al.* Separation of emission mechanisms in spintronic terahertz emitters. *Phys. Rev. B* **104**, 064419 (2021).

52 Rouzegar, R. *et al.* Laser-induced terahertz spin transport in magnetic nanostructures arises from the same force as ultrafast demagnetization. *Phys. Rev. B* **106**, 144427 (2022).

53 Xu, Y. *et al.* Inverse Orbital Hall Effect Discovered from Light-Induced Terahertz Emission. Prepint at https://arxiv.org/abs/2208.01866 (2022).

54 Choi, G.-M., Moon, C.-H., Min, B.-C., Lee, K.-J. & Cahill, D. G. Thermal spin-transfer torque driven by the spin-dependent Seebeck effect in metallic spin-valves. *Nat. Phys.* **11**, 576-581 (2015).

55 Alekhin, A. *et al.* Femtosecond Spin Current Pulses Generated by the Nonthermal Spin-Dependent Seebeck Effect and Interacting with Ferromagnets in Spin Valves. *Phys. Rev. Lett.* **119**, 017202 (2017).

56 Razdolski, I. *et al.* Nanoscale interface confinement of ultrafast spin transfer torque driving non-uniform spin dynamics. *Nat. Commun.* **8**, 15007 (2017).

57 Kimling, J. *et al.* Picosecond Spin Seebeck Effect. *Phys. Rev. Lett.* **118**, 057201 (2017).

58 Chen, Y.-J. & Huang, S.-Y. Light-induced thermal spin current. *Phys. Rev. B* **99**, 094426 (2019).

59 Bartell, J. M., Ngai, D. H., Leng, Z. & Fuchs, G. D. Towards a table-top microscope for nanoscale magnetic imaging using picosecond thermal gradients. *Nat. Commun.* **6**, 8460 (2015).

60 Seifert, T. S. *et al.* Femtosecond formation dynamics of the spin Seebeck effect revealed by terahertz spectroscopy. *Nat. Commun.* **9**, 2899 (2018).

61 Lu, W. *et al.* Ultrafast photothermoelectric effect in Dirac semimetallic $Cd_3As_2$ revealed by terahertz emission. *Nat. Commun.* **13**, 1623 (2022).

62 Chuang, T. C., Su, P. L., Wu, P. H. & Huang, S. Y. Enhancement of the anomalous Nernst effect in ferromagnetic thin films. *Phys. Rev. B* **96**, 174406 (2017).

63 Lombard, J., Detcheverry, F. & Merabia, S. Influence of the electron–phonon interfacial conductance on the thermal transport at metal/dielectric interfaces. *J. Phys.: Condens. Matter* **27**, 015007 (2015).

64 Querry, M. R. Optical constants. *Contractor Report*, CRDC-CR-85034 (1985).

65 Duan, Z. *et al.* The Longitudinal Spin Seebeck Coefficient of Fe. *IEEE Magn. Lett.* **10**, 1-5 (2019).

66 Tian, Y., Ye, L. & Jin, X. Proper Scaling of the Anomalous Hall Effect. *Phys. Rev. Lett.* **103**, 087206 (2009).

67 Kadlec, F., Kužel, P. & Coutaz, J.-L. Optical rectification at metal surfaces. *Opt. Lett.* **29**, 2674-2676 (2004).

68 Costa, J. D. *et al.* Terahertz dynamics of spins and charges in CoFe/$Al_2O_3$ multilayers. *Phys. Rev. B* **91**, 104407 (2015).

69 Guo, L., Hodson, S. L., Fisher, T. S. & Xu, X. Heat Transfer Across Metal-Dielectric Interfaces During Ultrafast-Laser Heating. *J. Heat Transfer* **134** (2012).

## Acknowledgments

We thank Chuanyun Tang and Xingliang Xu for their help on wire-bonding and AHE voltage measurement respectively. This work is supported by National Nature Science Foundation of China (NSFC) (Grants No. 62027807, No. 61988102, No. 62005256, No. 11974165, No. 61975110, No. 51971110, No. 52176075), and the National Key R&D Program of China (Grants No. 2021YFA1401400). Ssu-Yen Huang, Po-Hsun Wu and Yi-Jia Chen acknowledge the support from the Ministry of Science and Technology of Taiwan under Grant No. MOST 111-2123-M-002-010.

## Author contributions

Z. Feng, Z. M. Jin, B. F. Miao, and S.-Y. Huang conceived the idea, designed the experiments, and supervised the project. Z. Feng built the bidirectional-pump THz emission spectroscopy. W. Tan provided the separation method of different contributions. Y.-J. Chen, Z. F. Zhong, P. H. Wu, and J. Cheng prepared the samples. Z. Feng, Z. M. Jin, Y. X Jiang, D. C. Wang, Y. M. Zhu, and G. H. Ma carried out the THz emission, ANE voltages and AHE voltages measurements. L. Zhang, L. Guo, S. Sun, S. Shin, J. Tang performed ultrafast temperature calculations. Z. Feng, W. Tan, Z. M. Jin, L. Zhang, B. F. Miao, H. F. Ding, L. Guo, D. Z. Hou, and S.-Y. Huang analyzed and discussed the data. Z. Feng wrote the original draft, and W. Tan, Z. M. Jin, B. F. Miao, H. F. Ding, L. Guo, D. Z. Hou, S.-Y. Huang reviewed and revised the manuscript.

## Competing interests

The authors declare no competing interests.

## Additional information

**Supplementary information**

**Correspondence** and requests for materials should be addressed to Zheng Feng, Zuanming Jin, Bingfeng Miao or Ssu-Yen Huang.

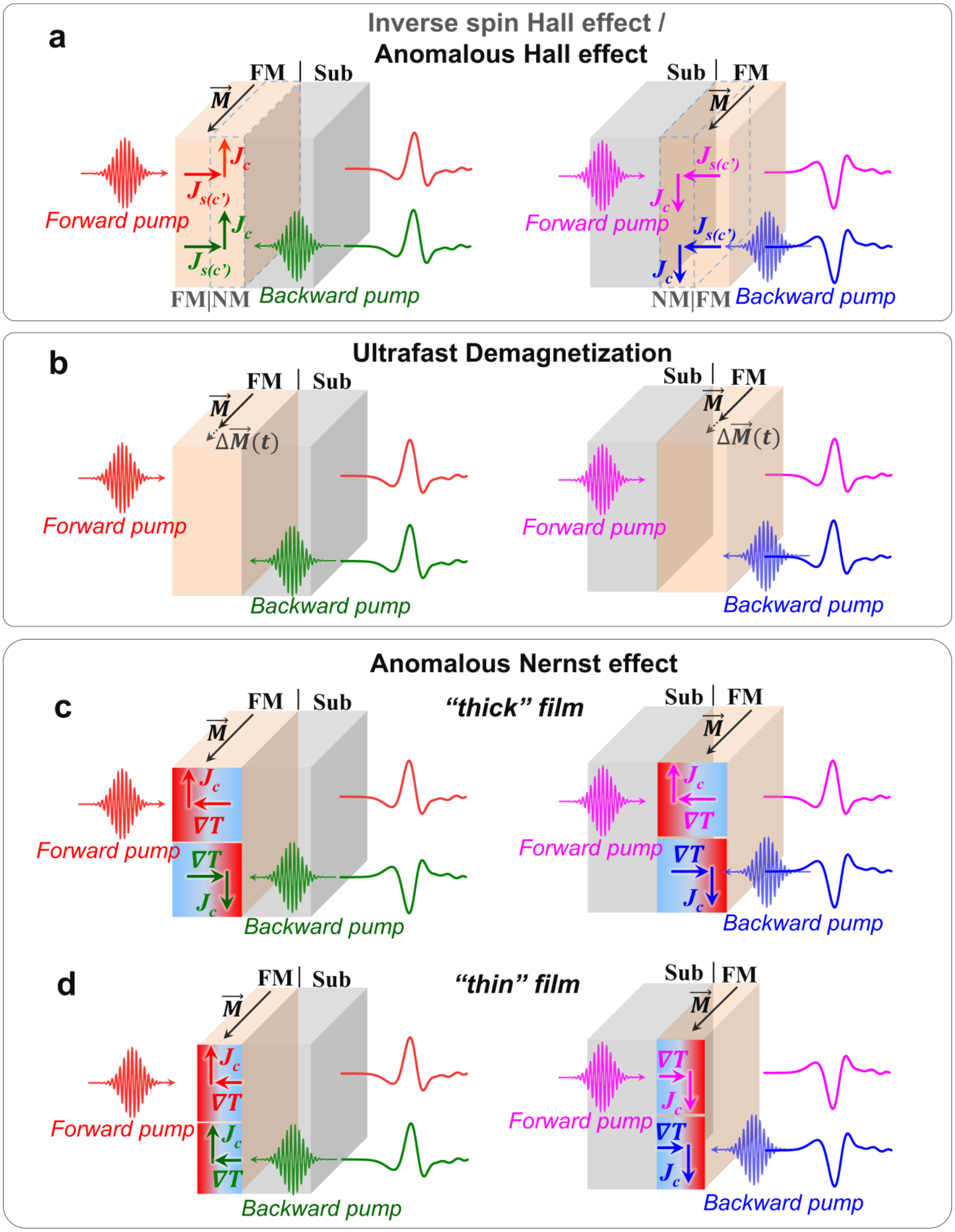

**Fig. 1 | Schematic of distinctive features of THz emission induced by different mechanisms.** (**a**) inverse spin Hall effect in FM|NM bilayer and anomalous Hall effect in single FM layer, (**b**) ultrafast demagnetization in single FM film, (**c**) anomalous Nernst effect in single "thick" FM film, and (**d**) in single "thin" FM film. The features are based on the analysis of THz signals under four different configurations, i.e., FM|Sub_forward pump (red), FM|Sub_backward pump (green), Sub|FM_forward pump (pink), Sub|FM_backward pump (blue), respectively.

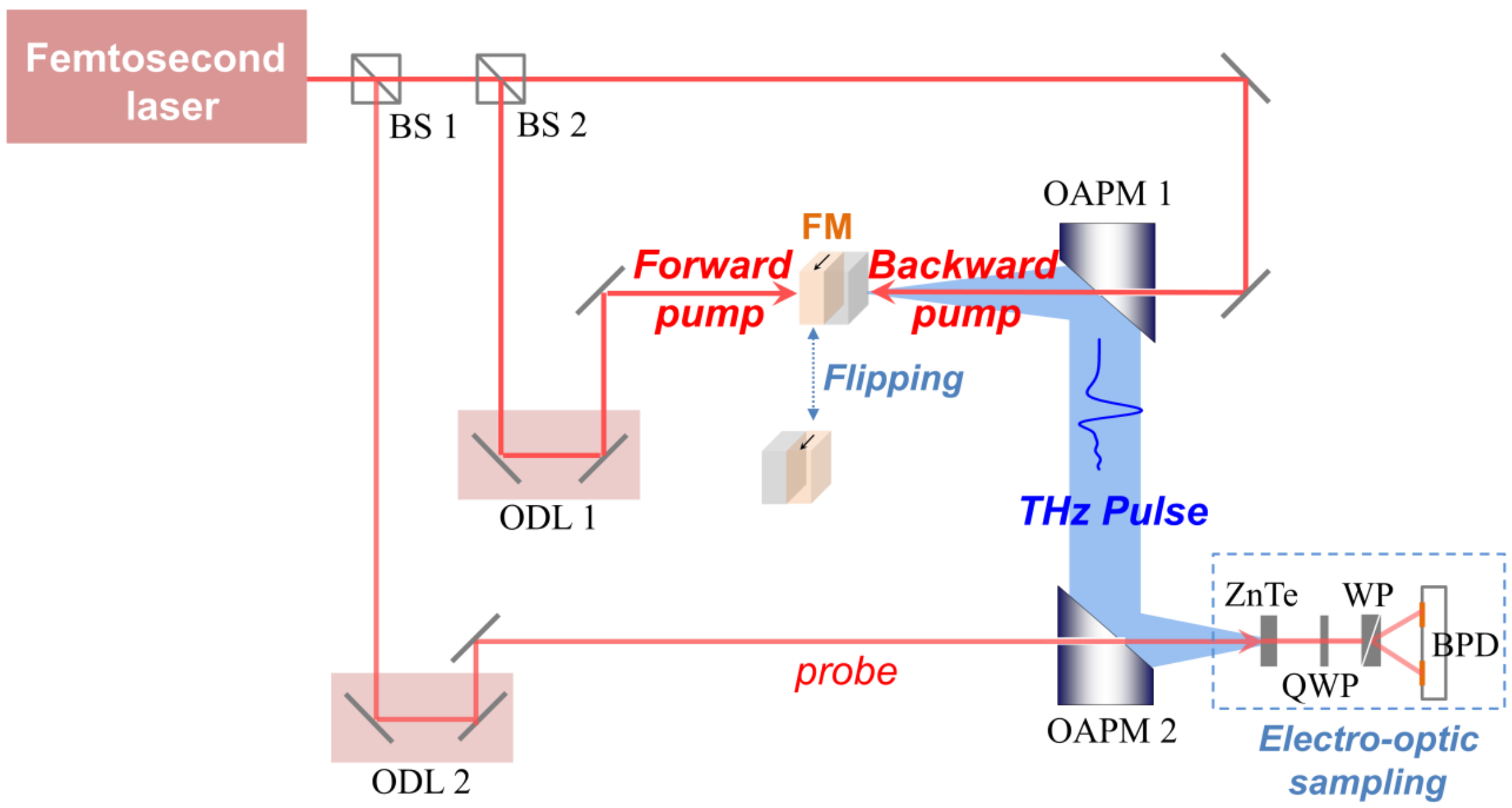

**Fig. 2 | Schematic of home-built bidirectional-pump THz emission spectroscopy.** In this setup, measurements can be taken by switching the laser pump direction and flipping the sample. BS: beam splitter. ODL: optical delay line, OAPM: off-axis parabolic mirror, QWP: quarter-wave plate, WP: Wollaston prism, BPD: balanced photodetector.

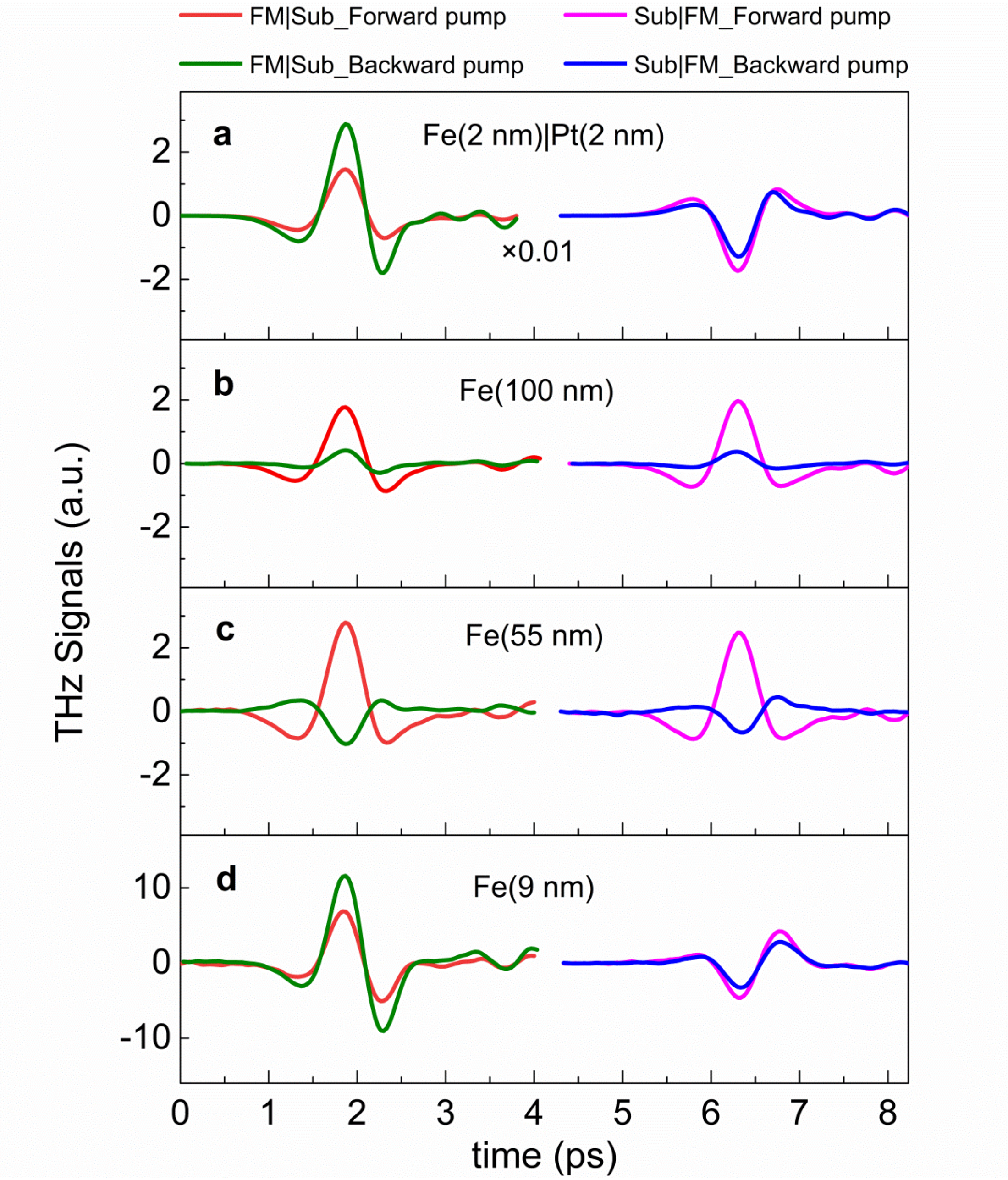

**Fig. 3 | Time-domain THz signals of different samples with four different configurations.** (**a**) Fe(2 nm)|Pt(2 nm), (**b**) Fe(100 nm), (**c**) Fe(55 nm), (**d**) Fe(9 nm). The typical THz amplitude of single Fe film is about two orders of magnitude smaller than that of Fe(2 nm)|Pt(2 nm).

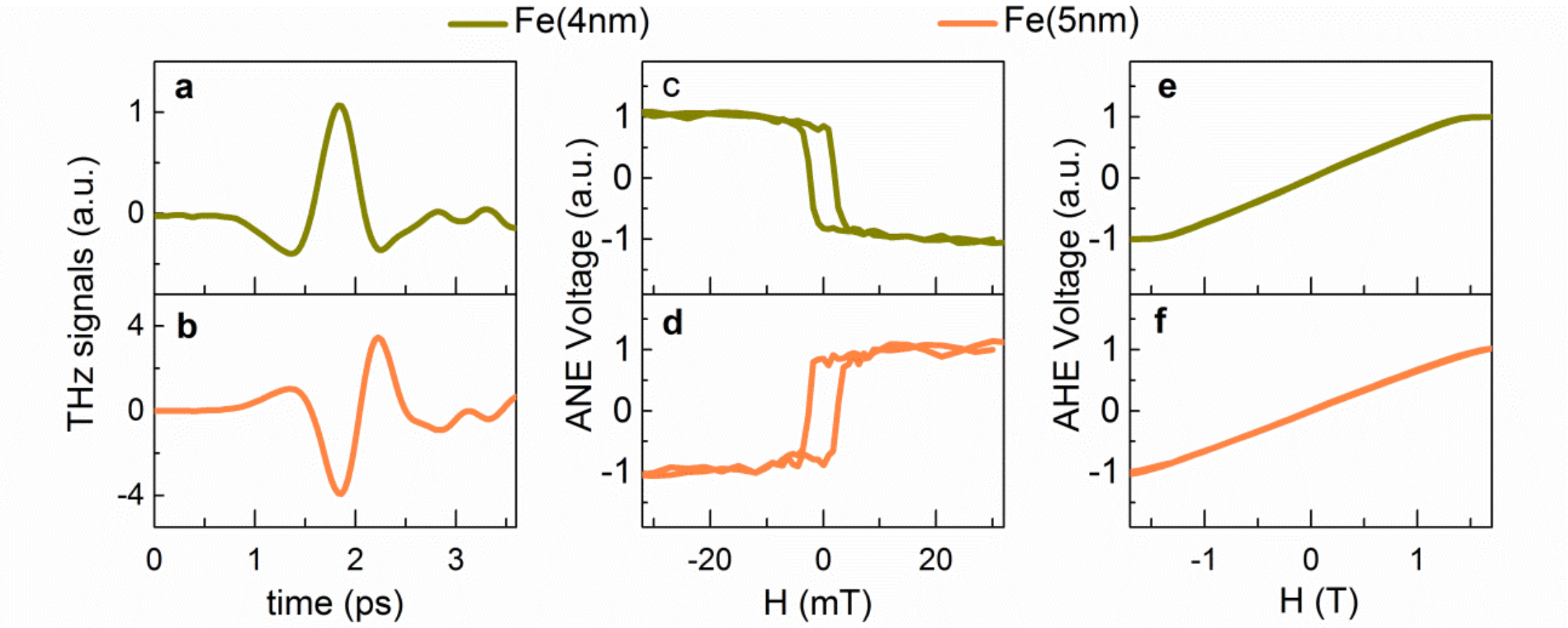

**Fig. 4 | THz signals and their corresponding ANE and AHE voltages of Fe(4 nm) and Fe(5 nm).** THz signals is in the Sub|FM_backward pump condition, while The ANE voltages were measured with the same laser pump condition. The AHE voltages were measured with the Van der Pauw method.

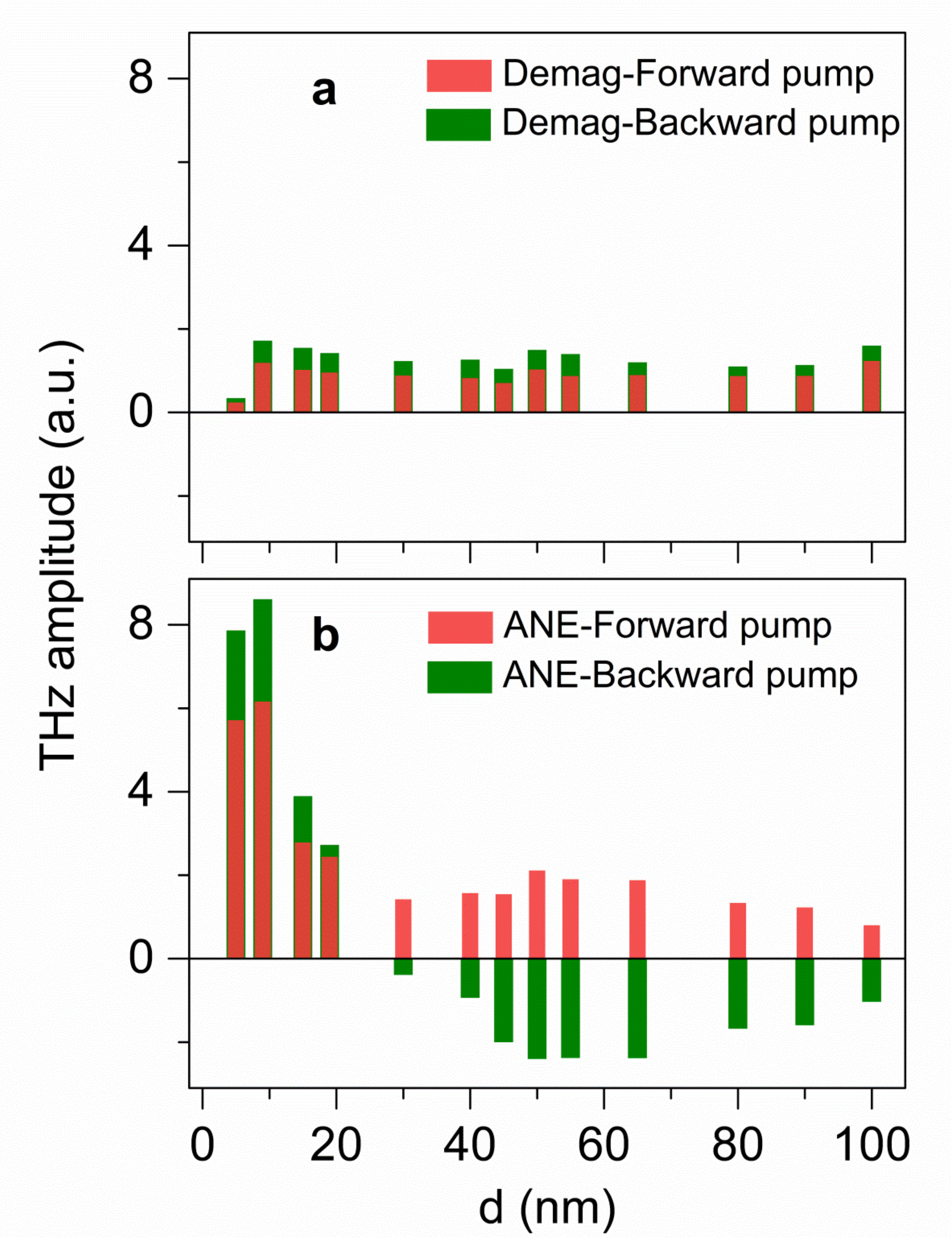

**Fig. 5 | The separated THz contribution as the Fe-thickness.** (**a**) Separated ultrafast demagnetization (Demag) induced THz amplitude, and (**b**) separated ANE induced THz amplitude, respectively.

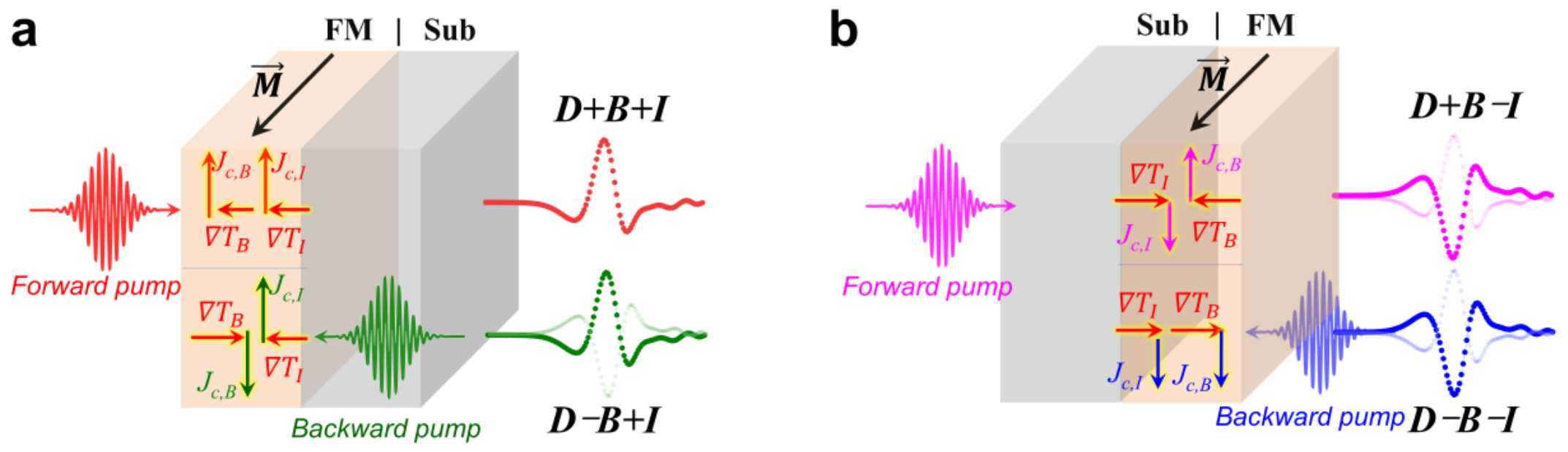

**Fig. 6 | Schematic of the separation method.** (**a**) The THz emission of FM|Sub under forwad and backward laser pump respectively. (**b**) The THz emission of Sub|FM under forwad and backward laser pump respectively. D: THz emission induced by ultrafast demagnetization. B: THz emission induced by ANE with "bulk" temperature gradient $\nabla T_B$. I: THz emission induced by ANE with "interface" temperature gradient $\nabla T_I$.

Supplementary information for

# Anomalous Nernst effect induced terahertz emission in a single ferromagnetic film

Zheng Feng[1]*, Wei Tan[1], Zuanming Jin[2]*, Yi-Jia Chen[3], Zhangfeng Zhong[4], Liang Zhang[5,7], Song Sun[1], Jin Tang[6], Yexing Jiang[2], Po-Hsun Wu[3], Jun Cheng[4], Bingfeng Miao[4,10]*, Haifeng Ding[4,10], Dacheng Wang[1], Yiming Zhu[2], Liang Guo[5], Sunmi Shin[7], Guohong Ma[8], Dazhi Hou[9,11], Ssu-Yen Huang[3]*

[1]Microsystem & Terahertz Research Center, CAEP, Chengdu 610200, P. R. China

[2]Terahertz Technology Innovation Research Institute, Terahertz Spectrum and Imaging Technology Cooperative Innovation Center, Shanghai Key Lab of Modern Optical System, University of Shanghai for Science and Technology, Shanghai 200093, P. R. China

[3]Department of Physics, National Taiwan University, Taipei 10617, Taiwan

[4]National Laboratory of Solid State Microstructures and Department of Physics, Nanjing University, Nanjing 210093, P. R. China

[5]Department of Mechanical and Energy Engineering, Southern University of Science and Technology, Shenzhen 518055, P. R. China

[6]School of Physics and Optoelectronics Engineering Science, Anhui University, Hefei 230601, P. R. China

[7]Department of Mechanical Engineering, National University of Singapore, 117516, Singapore

[8]Department of Physics, Shanghai University, Shanghai 200444, P. R. China.

[9]ICQD, Hefei National Laboratory for Physical Sciences at Microscale, University of Science and Technology of China, Hefei 230026, P. R. China

[10]Collaborative Innovation Center of Advanced Microstructures, Nanjing University, Nanjing 210093, P. R. China

[11]Department of Physics, University of Science and Technology of China, Hefei 230026, P. R. China

*fengzheng_mtrc@caep.cn, *physics_jzm@usst.edu.cn , *bfmiao@nju.edu.cn, *syhuang@phys.ntu.edu.tw

## Section S1. Ultrafast temperature upon femtosecond laser pulse illumination

When femtosecond (fs) laser pulse illuminates on a metal, the laser energy is first deposited into electrons instantaneously while the lattice remains cold. On a fs timescale, the energy is distributed among the electrons by electron-electron coupling, leading to thermalization of the electrons. Then the electrons transfer their energy to the lattice due to electron-phonon coupling, which occurs from <1 ps to a few tens of picoseconds. This can be described by the two-temperature model with the electron temperature $T_e$ and phonon temperature $T_p$ [1].

As the anomalous Nernst effect (ANE) is driven by the electron temperature gradient, we focus on the ultrafast electron temperature based on the two-temperature model. For a metal on a dielectric substrate, when we don't consider the heat exchange between the metal and the substrate, the electron temperature gradient along the film normal direction is determined by the laser absorption profile. As illustrated in Fig. S1a and Fig. S1b, the electron temperature $T_e$ (dashed line) at the side facing to the laser incident direction is always higher than the that at other side. When changing the incident direction of the laser, the electron temperature profile is reversed. However, as shown in Fig S1c, besides the electron-electron coupling and electron-phonon coupling in metal, energy transfer between the metal and the dielectric substrate also occurs. In general, there are two energy transfer channels [2,3]. One is an indirect transfer, in which the energy is first transferred from metal electron to metal phonon via electron-phonon coupling, and then transferred to substrate via the phonon-phonon coupling between metal phonons and substrate phonons. This indirect coupling is longer than 1 ps. The other one is the direct transfer via the direct coupling between metal electrons and substrate phonons. As a result, the metal electron temperature near the metal|substrate interface may be lower than that in the metal (in the condition of the uniform laser energy absorption), leading to an electron temperature gradient pointing away from the metal|substrate interface (Light sphere color corresponds to low temperature in Fig S1c). Importantly, as the timescale of the coupling between metal electrons and substrate phonons can be shorter than 1 ps [2], an ultrafast electron temperature gradient near the interface can be established within the timescale of 1ps. Therefore, the profile of ultrafast electron temperature $T_e$ is modified from the dashed line to the solid line, as illustrated in Fig. S1a and Fig. S1b. And the electron temperature $T_e$ can be regarded as the superposition of "bulk" electron temperature and "interface" electron temperature, where the "bulk" component is determined by the laser energy absorption profile in metal, and the "interface" component is caused by the heat loss across the interface. Meanwhile, the gradient of the electron temperature $\nabla T$ can be divide into the "bulk" electron temperature gradient $\nabla T_B$ and the "interface" electron temperature gradient $\nabla T_I$. $\nabla T_B$ always points to the pump direction of laser, and $\nabla T_I$ points away from the FM|substrate interface regardless of the pump direction, as illustrated in Fig. S1a and Fig. S1b.

Since the interface contribution decreases and the laser absorption profile becomes sharper as the metal thickness increases, $\nabla T_I$ is dominant in "thin" films, while $\nabla T_B$ is dominant in "thick" films. The critical thickness between "thick" and "thin" is highly related to the optical penetration depth of metal and thermal properties of the sample. From the above discussion, the total electron temperature gradient $\nabla T$ is different for "thin" and "thick" films. For "thin" films, $\nabla T$ points away from the FM|substrate interface regardless of the pump direction, while for thick films, $\nabla T$ always points to the pump direction of laser. And this is confirmed by the ultrafast temperature simulation based on the two-temperature model, which is detailedly presented in the Methods. The ultrafast electron temperature simulation results of Fe films with thickness of 55 nm and 9 nm were shown in Fig. S2. For $SiO_2$(5 nm)|Fe(55 nm)|MgO substrate with thickness much larger than Fe optical penetration depth (17.3 nm), both the temperature gradient under forward and backward pump point to the pump direction of fs laser, as illustrated in Figs. S2a and S2b. However, the temperature gradient of the thin sample $SiO_2$(5 nm)|Fe(9 nm)|MgO substrate points away from the Fe|substrate interface regardless of the selection of the pump directions, as

illustrated in Figs. S2c and S2d. And the timescale of the temperature gradient is in the sub-ps range, as shown in the right part of the figures. Note that the addition of the 5 nm thick $SiO_2$ capping layer has little influence on the modified electron temperature profile (as shown with solid line) in Fig. S1a and Fig. S1b, because $SiO_2$ is very a good thermal insulator.

The thermophysical parameters used in the ultrafast temperature simulation are listed in Table S1. In the Table, $C_e(T_e)$ and $k_e(T_e)$ are described as [4,5]:

$$C_e(T_e) = -184590.00526 + 1996.3795T_e - 0.8783T_e^2 + 1.55491\times10^{-4}T_e^3 + 1.76139\times10^{-9}T_e^4 - 2.48208\times10^{-12}T_e^5 \quad \text{(S1)}$$

$$k_e(T_e) = 50.2T_e / T_p \quad \text{(S2)}$$

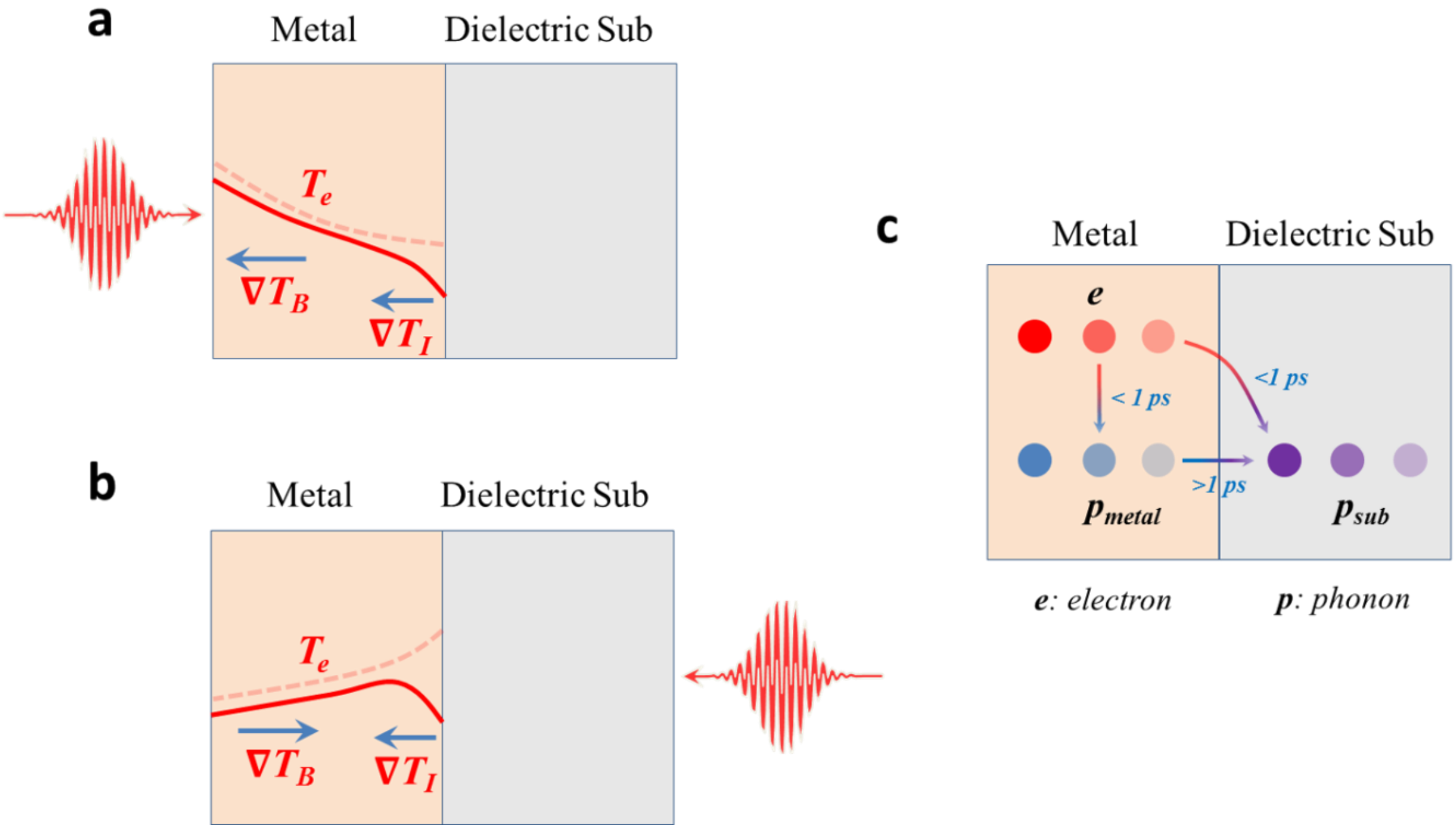

**Fig. S1 | Ultrafast electron temperature feature of Metal|Dielectric Substrate upon fs laser pulse illumination.** (a) Ultrafast electron temperature distribution with (solid line) and without (dashed line) coupling between metal electrons and substrate phonons with laser incident on the metal side. (b) Ultrafast electron temperature distribution with (solid line) and without (dashed line) coupling between metal electrons and substrate phonons with laser incident on the substrate side. (c) Schematic of the coupling between metal electrons and substrate phonons.

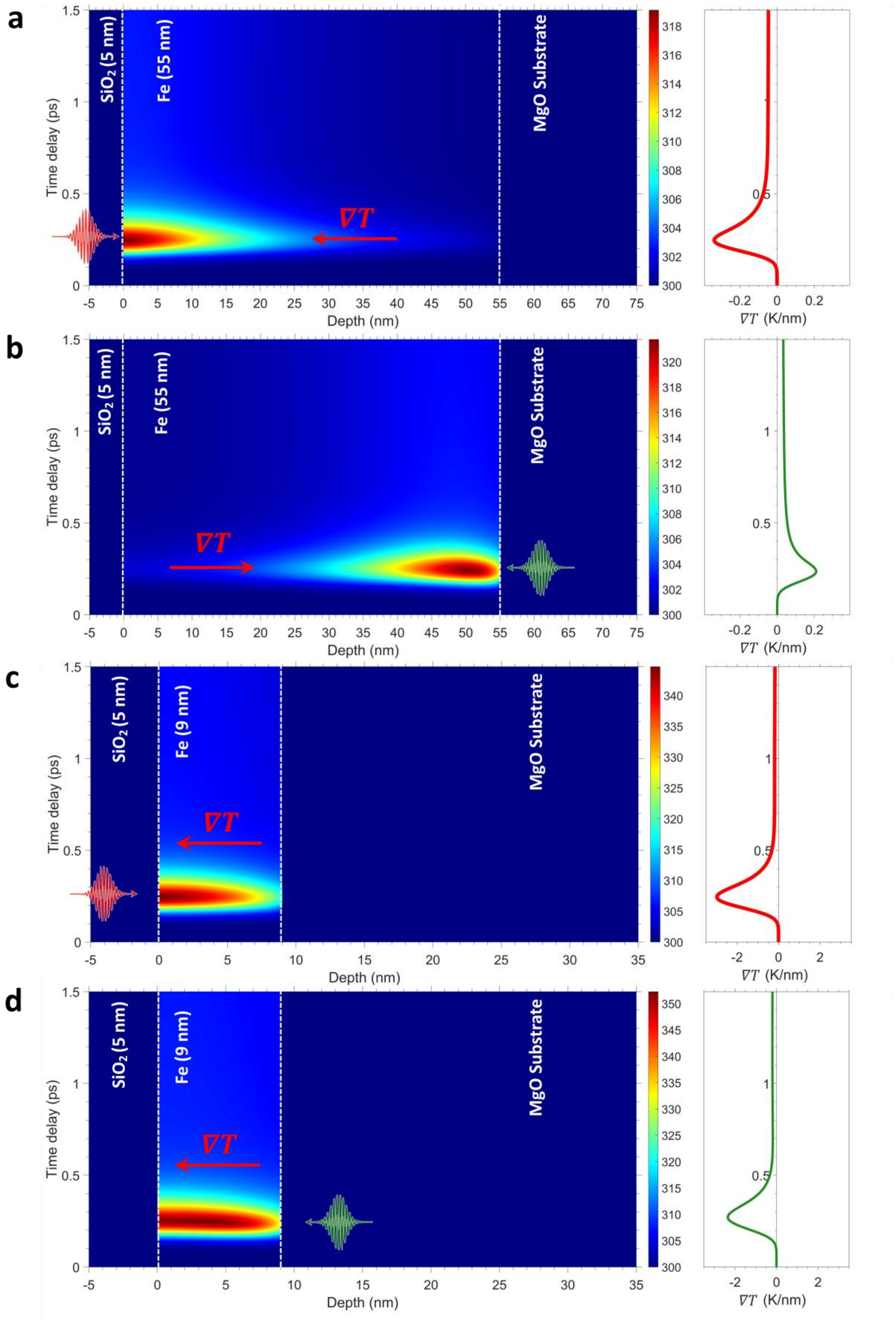

**Fig. S2 | The simulation results of ultrafast electron temperature distributions and temperature gradients.** (a) $SiO_2$(5 nm)|Fe(55 nm)|MgO substrate under forward pump. (b) $SiO_2$(5 nm)|Fe(55 nm)|MgO substrate under backward pump. (c) $SiO_2$(5 nm)|Fe(9 nm)|MgO substrate under forward pump. (d) $SiO_2$(5 nm)|Fe(55 nm)|MgO substrate under backward pump.

**Supplementary Table S1 | Main thermophysical parameters in the simulations**

| | | | | |
|---|---|---|---|---|
| **Parameter** | $C_e$ **(J·m$^{-3}$·K$^{-1}$)** | $C_p$ **(J·m$^{-3}$·K$^{-1}$)** | $C_c$ **(J·m$^{-3}$·K$^{-1}$)** | $C_s$ **(J·m$^{-3}$·K$^{-1}$)** |
| **Value** | $C_e(T_e)$ [4] | $3.32\times10^{6}$ [5] | $3.31\times10^{6}$ [6] | $3.06\times10^{6}$ [7] |
| **Parameter** | $k_e$ **(J·m$^{-3}$·K$^{-1}$)** | $k_p$ **(J·m$^{-3}$·K$^{-1}$)** | $k_c$ **(J·m$^{-3}$·K$^{-1}$)** | $k_s$ **(J·m$^{-3}$·K$^{-1}$)** |
| **Value** | $k_e(T_e)$ [5] | 9 [5] | 1.34 [6] | 59 [7] |
| **Parameter** | $R_{ec}$ **(K·m$^{2}$·W$^{-1}$)** | $R_{es}$ **(K·m$^{2}$·W$^{-1}$)** | $R_{pc}$ **(K·m$^{2}$·W$^{-1}$)** | $R_{ps}$ **(K·m$^{2}$·W$^{-1}$)** |
| **Value** | $10^{-7}$ | $3\times10^{-11}$ | $10^{-7}$ | $10^{-7}$ |
| **Parameter** | $G$ **(W·m$^{-3}$·K$^{-1}$)** | $\delta$ **(nm)** | $F$ **(J·m$^{-2}$)** | $t_p$ **(fs)** |
| **Value** | $32\times10^{17}$ [8] | 17.3 [9] | 0.7 | 100 |

## Section S2. The raw data of THz emission of Fe films with various thicknesses

Figure S3 shows the raw data of THz emission of Fe films with various thicknesses. We note that the differences of the shapes among the Fe films with different thicknesses mainly originate from the different light-matter interactions in the time domain, which are complicated and beyond the scope of our current manuscript. When we limit our discussion on the four measured signals of a specific Fe film, their shapes are either “dip-peak-dip” or “peak-dip-peak”. With a simple sign reversal operation, their shapes are actually very similar. The main difference lies in the relative amplitude of the peak and dip. The THz signal is the superposition of the components induced by different mechanisms which have very different symmetries. Therefore, it is intuitive and effective to inspect the sign/polarity of the THz pulses of an individual Fe film to distinguish different mechanisms based on their very different symmetries.

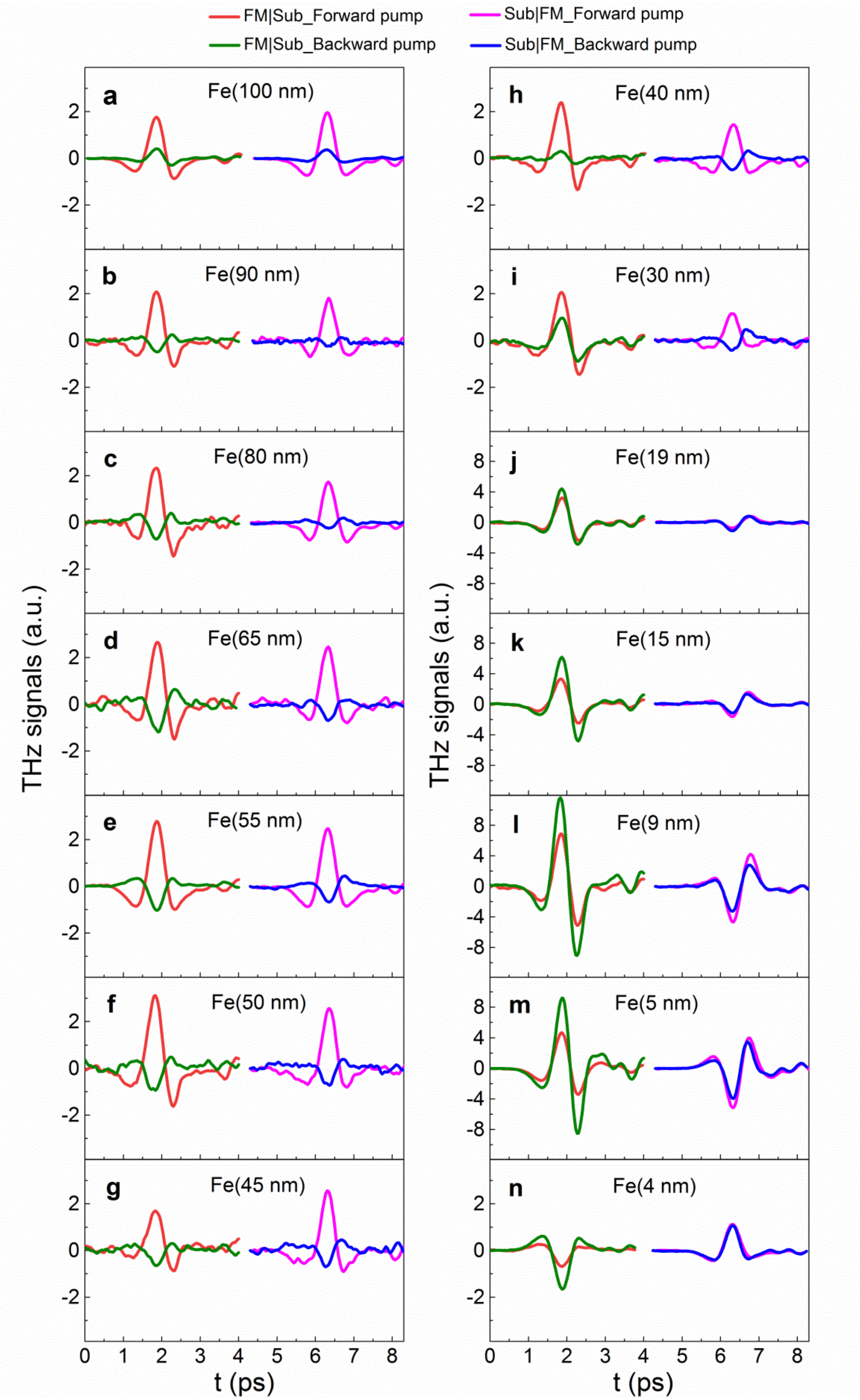

**Fig. S3 | The raw data of THz emission of Fe films with various thickness.**

## Section S3. The laser polarization dependence of THz emission

The nonlinear optical rectification effects in metals can lead to THz emission, which have been observed in normal metals when pumped with high density fs laser [10]. In principle, the optical rectification effects in FMs could also induce THz emission. Owning to their very different mechanisms, the THz emission based on the optical rectification effects and spin-related effects can be distinguished via the laser polarization dependent measurements. The physical principle lies in the facts that (i) the orientation of the emitted electric field based on the spin-related effects is perpendicular to the applied magnetic field but robust to the laser polarization, and (ii) the optical rectification effects commonly depend on the laser polarization [10,11]. Consequently, if we rotate the laser polarization while keep the sample and the applied magnetic field fixed [12,13], we can find the transparent evidence.

The measured results are depicted in Fig. S4. We firstly measured the laser polarization dependent THz signal from a ZnTe(110) crystal (a typical optical rectification effect based THz emitter), as shown in fig. S4a. It is clearly seen that the peak amplitude varies as a sine squared function, which presents a perfect agreement with the theoretical result of the traditional optical rectification effect [14]. In comparison, the THz signals from Fe films with different thickness (100 nm, 55 nm, and 9 nm) are independent of laser polarization. These results evidently exclude the role of the traditional optical rectification effects in Fe films. Similar method and conclusion have been also reported in CoFe nanofilms [12].

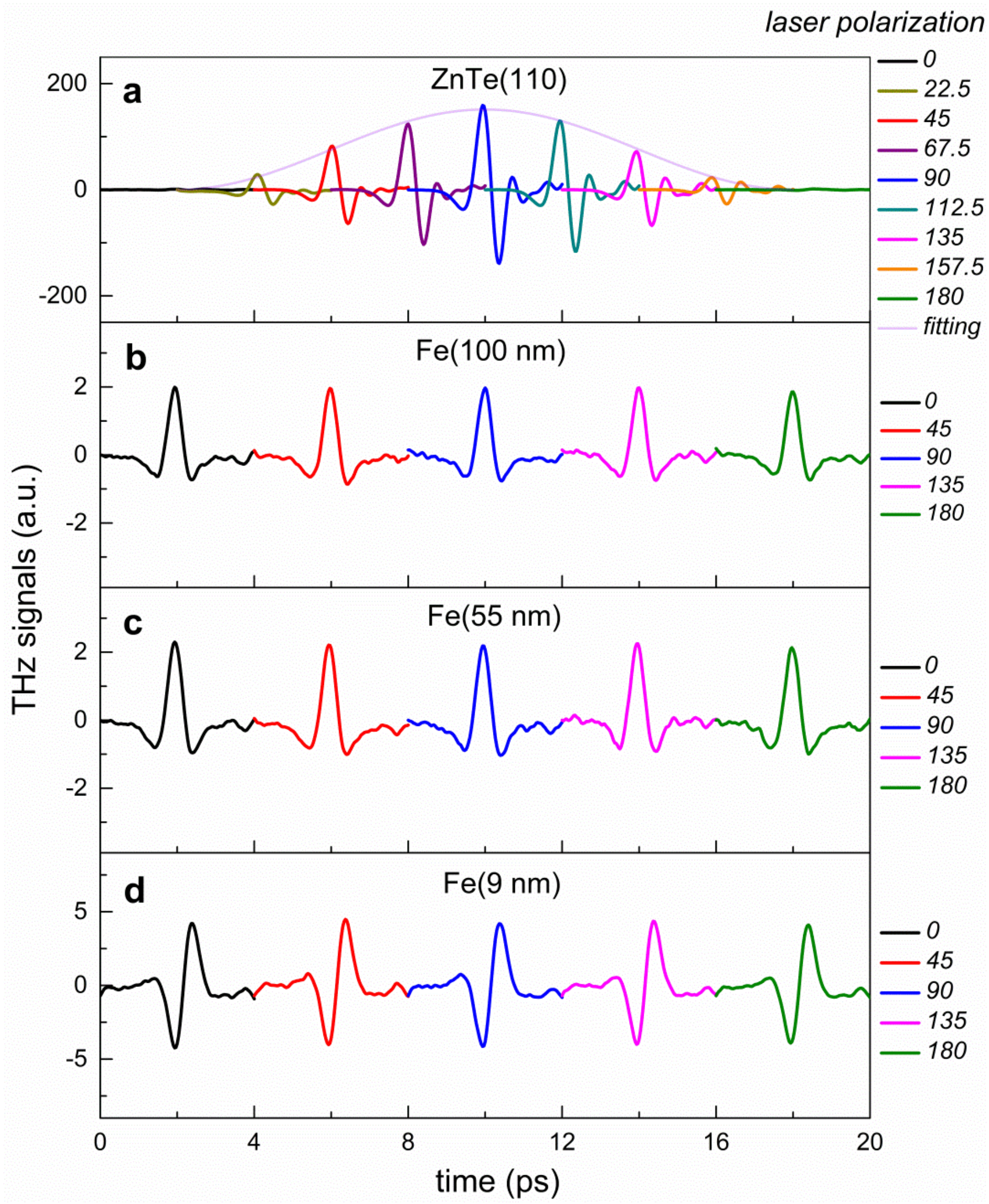

**Fig. S4 | Laser polarization dependence of THz signals from different samples.** (a) ZnTe (100) crystal. (b) Fe (100 nm). (c)Fe (55 nm). (d)Fe (9 nm).

## Section S4. Separation of the THz emission induced by the ultrafast demagnetization and ANE

Figure S5(a1-e1) show the four THz waveforms which are normalized according to the method discussed in Methods and shifted to the same time position respectively, where the Fe thickness are chosen with five typical thicknesses (100 nm, 85 nm, 55 nm, 40 nm and 9 nm). Then the THz waveforms of Demagnetization($D$), ANE-Bulk($B$), and ANE-Interface($I$) were calculated and separated from the four THz waveforms according to Eqs. (19-21), and the separated results were shown in Figs. S5(a2-a4), Figs. S5(b2-b4), Figs. S5(c2-c4), Figs. S5(d2-d4) and Figs. S6(e2-e4) respectively. The yellow and black waveforms are obtained with two independent calculations, which are almost identical for every component with the same Fe thickness. This confirms the validity of our separation method and evidences that our method is self-consistent.

Figure S5 clearly shows that (i) when the Fe film is thin enough (9 nm), the contribution of ANE of "interface" temperature gradient unambiguously dominates; (ii) as the thickness increases into 40 nm, the contribution of ANE of "bulk" temperature gradient becomes more important; (iii) when the thickness reaches 100 nm, ultrafast demagnetization induced contribution becomes dominant.

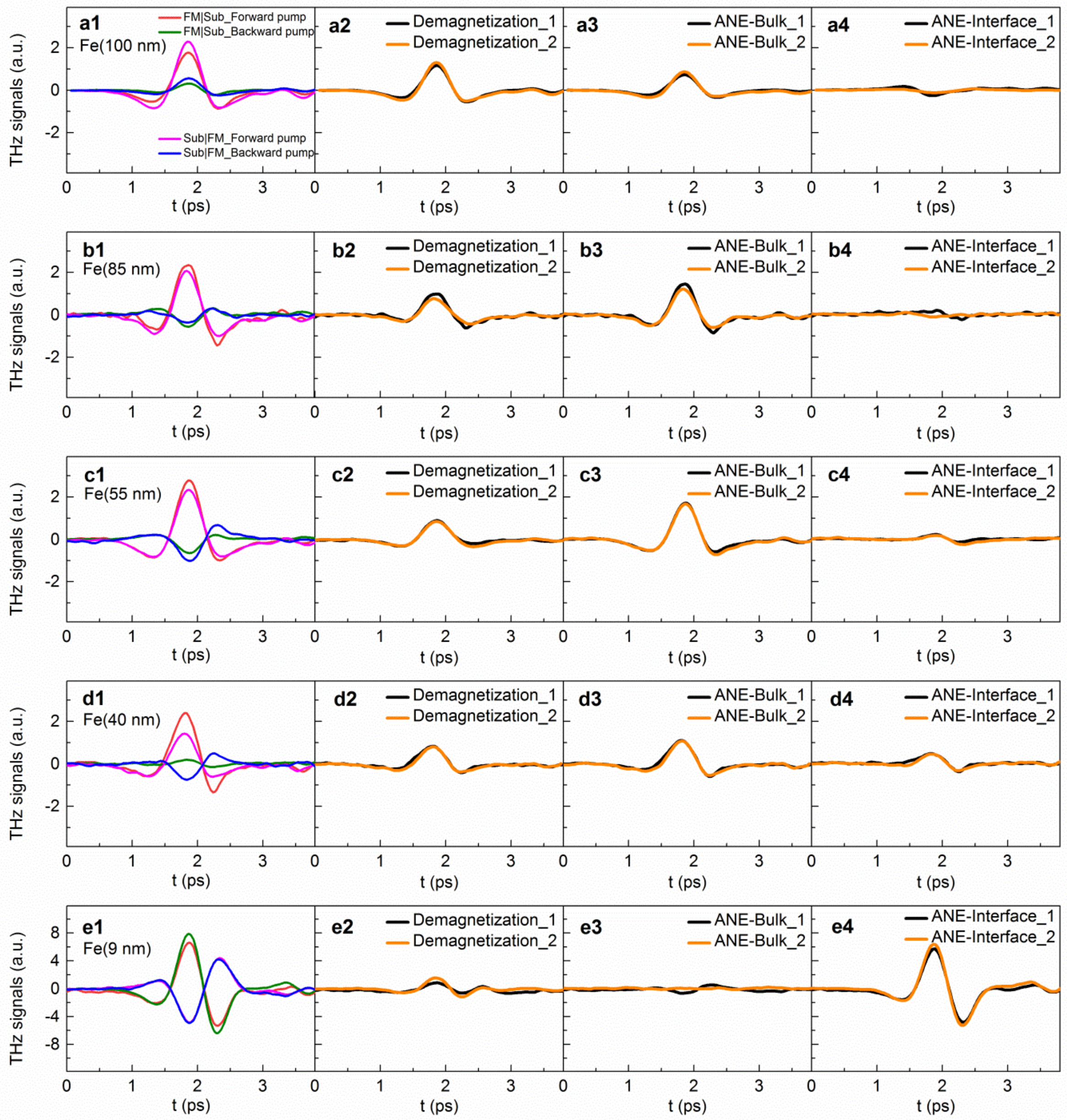

**Fig. S5 | The experimentally measured THz emission(normalized) and their separated components with two individual methods with Fe films with typical thicknesses.** (a1-a4) for Fe(100 nm), (b1-b4) for Fe(85 nm), (c1-c4) for Fe(55 nm), (d1-d4) for Fe(40 nm) and (e1-e4) for Fe(9 nm), respectively.

## Supplementary References

1 Eesley, G. L. Generation of nonequilibrium electron and lattice temperatures in copper by picosecond laser pulses. *Phys Rev. B* **33**, 2144-2151 (1986).

2 Lombard, J., Detcheverry, F. & Merabia, S. Influence of the electron–phonon interfacial conductance on the thermal transport at metal/dielectric interfaces. *J. Phys.: Condens. Matter* **27**, 015007 (2015).

3 Guo, L., Hodson, S. L., Fisher, T. S. & Xu, X. Heat Transfer Across Metal-Dielectric Interfaces During Ultrafast-Laser Heating. *J. Heat Transfer* **134** (2012).

4 Zhigilei, Z. L. a. L. V. Electron-phonon coupling and electron heat capacity in metals at high electron temperatures, https://compmat.org/electron-phonon-coupling/.

5 Kang, K. & Choi, G.-M. Electron-Phonon Coupling Parameter of Ferromagnetic Metal Fe and Co. *Materials* **14**, 2755 (2021).

6 Andersson, S. & Dzhavadov, L. Thermal conductivity and heat capacity of amorphous SiO2: pressure and volume dependence. *J. Phys.: Condens. Matter* **4**, 6209 (1992).

7 https://www.korth.de/en/materials/detail/Magnesium%20Oxide.

8 Simoni, J. & Daligault, J. First-Principles Determination of Electron-Ion Couplings in the Warm Dense Matter Regime. *Phys. Rev. Lett.* **122**, 205001 (2019).

9 Querry, M. R. Optical constants. *Contractor Report*, CRDC-CR-85034 (1985).

10 Kadlec, F., Kužel, P. & Coutaz, J.-L. Optical rectification at metal surfaces. *Opt. Lett.* **29**, 2674-2676 (2004).

11 Rice, A. *et al.* Terahertz optical rectification from 〈110〉 zinc-blende crystals. *Appl. Phys. Lett.* **64**, 1324-1326 (1994).

12 Costa, J. D. *et al.* Terahertz dynamics of spins and charges in CoFe/Al2O3 multilayers. *Phys. Rev. B* **91**, 104407 (2015).

13 Cheng, L. *et al.* Far out-of-equilibrium spin populations trigger giant spin injection into atomically thin MoS2. *Nat. Phys.* **15**, 347-351 (2019).

14 Blanchard, F. *et al.* Generation of 1.5 µJ single-cycle terahertz pulses by optical rectification from a large aperture ZnTe crystal. *Opt. Express* **15**, 13212-13220 (2007).